# Utilizing Model-Free Reinforcement Learning for Optimizing Secure Multi-Party Computation Protocols

1st Javad Sayyadi
Faculty of Electrical & Computer Engineering
University of Tabriz
Tabriz, Iran
J.sayyadi1400@ms.tabrizu.ac.ir

2nd Mahdi Nangir
Faculty of Electrical & Computer Engineering
University of Tabriz
Tabriz, Iran
Nangir@tabrizu.ac.ir

3rd Mahmood Mohassel Feghhi
Faculty of Electrical & Computer Engineering
University of Tabriz
Tabriz, Iran
Mohasselfeghhi@tabrizu.ac.ir

4th Hamid Sayyadi
Faculty of Electrical & Computer Engineering
University of Tabriz
Tabriz, Iran
H.sayyadi1400@ms.tabrizu.ac.ir

*Abstract*—In this manuscript, we explore the application of model-free reinforcement learning in optimizing secure multiparty computation (SMPC) protocols. SMPC is a crucial tool for performing computations on private data without the need to disclose it, holding significant importance in various domains, including information security and privacy. However, the efficiency of current protocols is often suboptimal due to computational and communicational complexities. Our proposed approach leverages model-free reinforcement learning algorithms to enhance the performance of these protocols. We have designed a reinforcement learning model capable of dynamically learning and adapting optimal strategies for secure computations. Our experimental results demonstrate that employing this method leads to a substantial reduction in execution time and communication costs of the protocols. These achievements highlight the high potential of reinforcement learning in improving the efficiency of secure multiparty computation protocols, providing an effective solution to the existing challenges in this field.

## I. Introduction

In today's world, where data increasingly plays a critical role in various decision-making processes and analyses, maintaining the security and privacy of this data is of paramount importance. Secure multiparty computation (SMPC) emerges as a powerful method for performing computations on private data without needing to disclose it, gaining significant traction in diverse fields such as information security, medicine, finance, and other sensitive domains. However, current protocols often face numerous challenges that can impact their efficiency and reliability.

One of the major issues with SMPC protocols is their computational complexity and high communication requirements, leading to increased execution time and resource consumption. This problem becomes particularly pronounced in distributed environments with numerous participating parties and large data volumes. Additionally, many existing protocols struggle to adapt to dynamic conditions and sudden changes in the computational environment, which can affect the overall security and efficiency of the system [1], [2].

The aim of this research is to investigate and optimize SMPC protocols using model-free reinforcement learning. We believe that reinforcement learning algorithms, with their capabilities in learning and optimizing dynamic strategies, can effectively address the existing problems and challenges in SMPC protocols. Driving Force for This investigation arises due to the need for efficient and reliable solutions to perform secure computations in distributed environments with large datasets. Many current SMPC protocols cannot adapt to dynamic conditions and sudden changes in the environment. This research seeks to provide a solution that can dynamically adapt to environmental changes while maintaining optimal performance [3].

We put forward a reinforcement learning model in this work that can dynamically learn and adapt optimal strategies for executing SMPC protocols. Experimental results indicate a notable reduction during run time and communication costs of the protocols, demonstrating the high potential of this method in improving the efficiency of secure multiparty computations.

## II. BACKGROUND AND RELATED WORK

Reinforcement learning (RL) is a subset of machine learning where an agent learns to make optimal decisions through interactions with the environment, trial and error, and receiving rewards. The goal is to maximize the cumulative reward. RL algorithms are divided into two categories: model-based and model-free. In model-free methods, the agent learns optimal policies based solely on past experiences and received rewards without having an accurate model of the environment. This characteristic makes model-free reinforcement learning suitable for complex and unpredictable problems [4], [5].

Secure multiparty computation (SMPC) refers to a set of protocols and techniques that allow multiple parties to jointly and securely perform computations on their private data without revealing their private information to others [6]. These protocols are used in various applications such as electronic voting, medical data sharing, and financial data analysis. The main challenge in SMPC is maintaining the security and privacy of the data while improving computational and communication efficiency [7].

Article [8], using deep reinforcement learning (DRL) algorithms, has achieved significant improvements in reducing costs, increasing the quality of service (QoS), and reducing response time in software-defined secure middle platforms. These results indicate that the proposed algorithm has effectively managed load balancing and provided better performance compared to existing methods, demonstrating the high potential and capability of DRL in optimizing costs and resources. One of the weaknesses of this article could be the complexity and computational resources required to implement DRL algorithms. Implementing these algorithms typically requires a high level of expertise in machine learning and advanced computational resources, which might be challenging in real-world environments with limited resources [9], [10].

Article [11] specifically examines the use of SMPC in energy consumption flexibility markets, addressing the challenges of data privacy and security in these systems. This paper explores methods where consumers submit encrypted bids to the market, helping to optimize the balance of supply and demand in exchange for financial compensation. Additionally, the paper demonstrates that the use of double auctions and encryption-based models can enhance the resource allocation process and improve security [12].

Article [13] addresses enhancing privacy in cloud environments. This paper highlights the importance of protecting sensitive data in cloud networks and introduces SMPC techniques as a key method for performing computations on encrypted data without revealing users' private information. The proposed methods in this research include advanced cryptographic algorithms and communication protocols for effectively implementing SMPC in cloud environments. The paper clearly demonstrates that SMPC can enhance privacy in cloud systems and lays the groundwork for the use of more advanced techniques for data protection and cloud network security [14].

Article [15] focuses on training and inference of machine learning models on vertically partitioned data. This paper addresses the issues and challenges associated with horizontally partitioned data, where different features of the data are distributed across various organizations, and proposes methods for performing computations on encrypted data. By using MPC, the paper demonstrates how training and inference processes of machine learning models can be conducted effectively and confidentially without disclosing sensitive information. The research clearly emphasizes the importance of data privacy protection in multi-party environments and introduces new techniques for the effective implementation of MPC in complex data scenarios [16].

Article [17] examines the challenges of interactive decision-making in reinforcement learning with value function approximation. This paper, using the Decision-Estimation Coefficient (DEC) framework—a statistical complexity measure for determining optimal regret bounds in interactive decision-making—aims to improve existing algorithms. By combining the Estimation-to-Decision method and optimistic estimation, the authors have achieved stronger theoretical results that could enhance the performance of reinforcement learning algorithms in complex environments. This research is theoretically significant as it provides a structured approach to improving the interaction between learning algorithms and decision-making processes, leading to better performance of model-free RL methods. However, one of the main challenges is the practical implementation and validation of results in real-world scenarios, which requires further investigation and adaptation [18].

## III. MODEL-FREE REINFORCEMENT LEARNING METHOD AND ITS APPLICATION TO SECURE MULTIPARTY COMPUTATION PROTOCOLS

In this section, we present a model-free RL method for optimizing SMPC protocols.

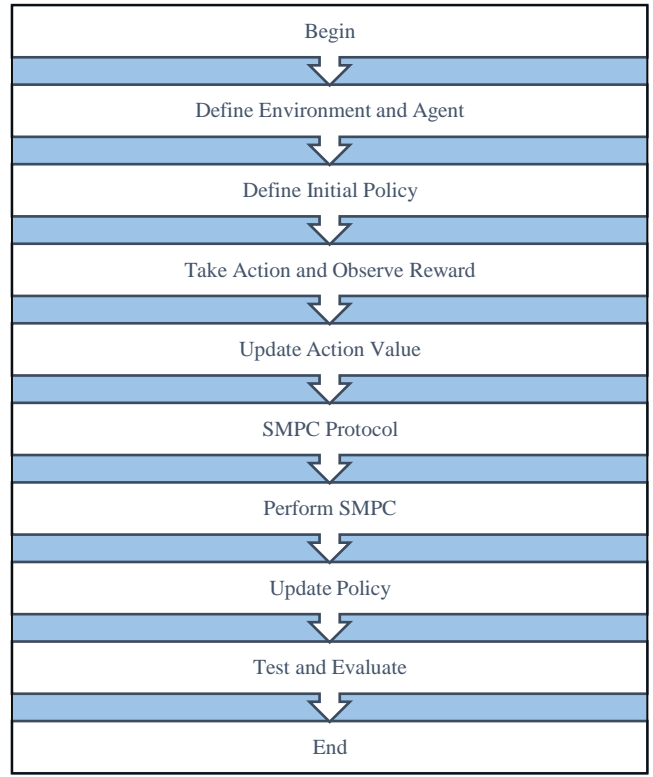

Fig. 1. *Model-Free Reinforcement Learning Method and Its Application to Secure Multiparty Computation Protocols*

Model-free reinforcement learning means that the algorithm does not require an explicit model of the environment and can perform optimization through interaction with the environment and collection of experiences. We utilize the Q-learning algorithm, which is a popular and efficient method in model-free reinforcement learning.

### A. General Framework of the Proposed Method

- Agent: Responsible for learning and optimizing the SMPC protocol.
- Environment: The SMPC protocol, which includes various parameters that need to be optimized.
- States: Represent the current settings of the SMPC protocol parameters.
- Actions: Changes in the parameters of the SMPC protocol.
- Rewards: Feedback received by the agent based on the performance and security of the protocol.

### B. Applying RL to SMPC Protocols

The RL agent begins by simulating the SMPC environment. At each step, the agent selects an action that represents the adjustment of one or more parameters of the SMPC protocol. The agent then receives a reward based on the execution of the protocol and the observed results. This reward is calculated based on performance metrics (such as execution time, computational cost) and security metrics (such as information security bit).

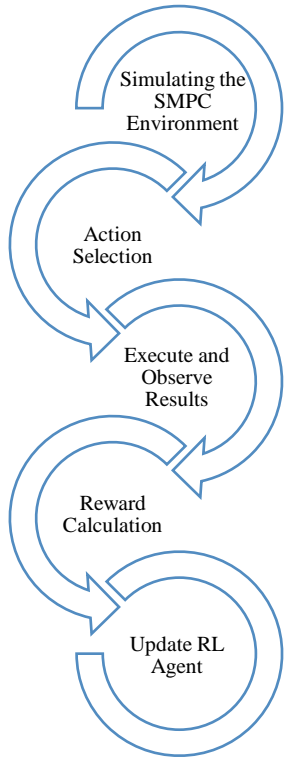

Fig. 2. *Applying RL to SMPC Protocols*

The first step involves starting with the simulation of the SMPC environment, representing the relevant protocols. In the second step, the reinforcement learning (RL) agent uses reinforcement learning policies to select and adjust one or more parameters of the SMPC protocol. In the third step, the SMPC protocol is executed with the parameters set by the RL agent, and the results, such as execution time and security bit, are observed. The fourth step includes calculating the reward based on performance and security criteria, where performance criteria include execution time and computational cost, and security criteria include the bit of information security. Finally, in the fifth step, the RL agent updates its Q-table or model using the received reward to make better decisions in subsequent stages. This process is iteratively repeated until the optimal settings for the SMPC protocol are achieved.

## IV. PROBLEM MODELING

We model the optimization of SMPC protocol parameters as a reinforcement learning problem. In this model, states (S) represent the current settings of the SMPC protocol, including values for various parameters such as the number of iteration rounds(of course, it is worth mentioning that SMPC protocols are **usually** completed in a single computational round.), data block sizes, and security bits. Actions (A) in this model indicate changes to the protocol parameters, such as increasing or decreasing the number of iteration rounds. Rewards (R) are calculated based on the protocol's performance, which includes performance metrics like execution time and computational cost, as well as security metrics such as the bit of information security. This reinforcement learning model allows us to achieve optimal settings for executing the SMPC protocol by using past experiences and updating protocol parameters.

### A. Definition of Key Parameters

The SMPC protocol includes key parameters whose optimization is crucial for achieving optimal performance. These parameters are: the number of iteration rounds and the size of data blocks. The number of iteration rounds refers to how many times the protocol needs to be executed to reach the final result. This parameter has a direct impact on computation time and the overall efficiency of the protocol. On the other hand, the size of data blocks determines the volume of data processed at each instance. Optimizing the size of data blocks can significantly impact processing speed and reduce communication costs. Consequently, optimal adjustment of these parameters can enhance both the efficiency and security of the SMPC protocol [20].

## V. IMPLEMENTATION OF THE Q-LEARNING ALGORITHM

We apply the Q-learning algorithm, which is a model-free RL method that fails to demand an explicit model of the environment. This algorithm operates using a Q-table that stores Q-values for each state-action pair. The Q-learning update formula is as follows:

$$[Q(s,a)] - \gamma max_{ta} Q(s',a') + \alpha[r + Q(s,a)] \quad (1)$$

where:

- $Q(s,a)$: The Q-value for state $s$ and action $a$.
- $\alpha$: Learning rate.
- $r$: Reward received after taking action $a$ in state $s$.
- $\gamma$: Discount factor.
- $s'$: New state after taking action $a$.

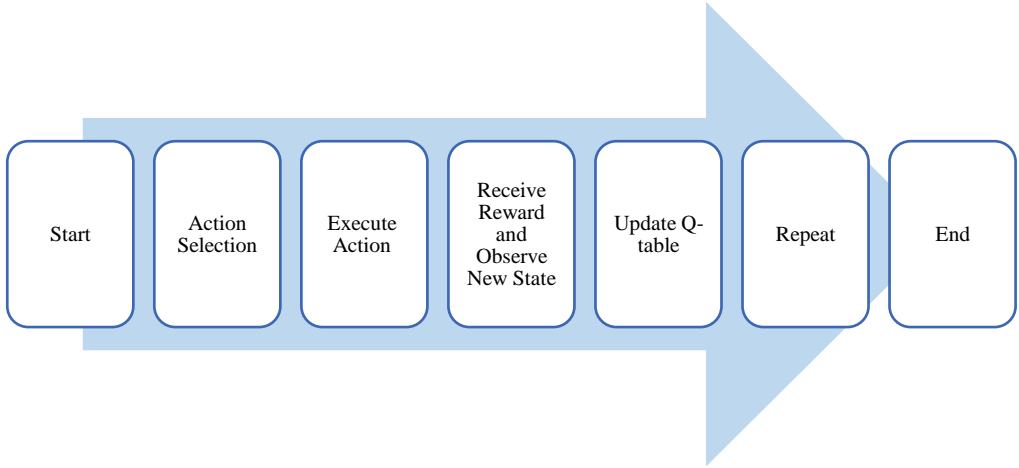

Fig. 3. *Implementation of the Q-learning Algorithm*

We need to show that this formula eventually converges to the optimal Q-value. The convergence proof of Q-learning is based on machine learning theories and Markov processes.

Objective Definition: The goal of Q-learning is to estimate the Q-value function such that, for all states and actions, $Q(s,a)$ converges to its optimal value $Q^*(s,a)$. $Q^*(s,a)$ is defined as follows:

$$E[r + \gamma max_{,a}Q^{(s',a')}|s,a] = Q^{(s',a')} \quad (2)$$

Q Update as a Weighted Average: Using the update formula, each update of the Q-value is obtained as a weighted average of the previous value and the new value. This weighted average is given by:

$$\alpha[r + \gamma max_{,a}Q(s',a')] + \alpha)Q(s,a) - 1) \leftarrow Q(s,a) \quad (3)$$

Convergence to the Optimal Value: Since $\alpha$ is a learning rate that typically decreases over time and approaches zero, this update formula ensures that the Q-value converges to the optimal value $Q^*$ with multiple iterations, then:

$$Q^*(s,a) \rightarrow Q(s,a) \quad (4)$$

Therefore, the Q-learning update formula is an effective method for estimating the optimal Q-value function in a Markov process. This formula, by combining the current Q-value with the reward return and future value estimation as a weighted average, converges to the optimal value over time.

The initialization of the Q-table starts with either zero or random values. Then, action selection is performed using the greedy $-\epsilon$ policy, where with probability $\epsilon$ a random action (exploration) is chosen, and with probability $\epsilon - 1$ the best action (exploitation) is selected. After executing the chosen action, the new state is observed and a reward is received. The Q-table is updated using the update formula:

$$[Q(s,a)] - \gamma max_{,a}Q(s',a') + \alpha[r + Q(s,a)] \quad (5)$$

These steps, from action selection to Q-table updating, are repeated until convergence or until a specified number of iterations is reached. Ultimately, the trained Q-table is used to make optimal decisions in each state.

### A. Exploitation Policies

Ensure that figures and tables are positioned at the top or bottom of columns and avoid placing them in the column's middle section. Large figures and tables may extend across both columns. Figure captions should be placed below the respective figures, and table headings should be positioned above the tables. Figures and tables should follow their citation in the text. Use the abbreviation "Fig. 1," even at the start of a sentence.

Action selection is performed using the greedy $-\epsilon$ and Soft max policies. In the greedy $-\epsilon$ policy, with probability $\epsilon$, a random action is chosen to encourage exploration, while with probability $\epsilon - 1$, the action with the highest Q-value is selected, which is known as exploitation. On the other hand, in the Soft max policy, the probability of selecting each action is calculated relative to the Q-value of that action. In this approach, actions with higher Q-values have a greater probability of being selected, but all actions still have a chance of being chosen, thus maintaining a balance between exploration and exploitation. These two policies are effectively used to enhance the efficiency and performance of RL algorithms.

### B. Training Phases

In the Q-learning process for optimizing SMPC protocol parameters, initial parameters such as the learning rate ($\alpha$), discount factor ($\gamma$), and exploration probability ($\epsilon$) are first set. Then, during the training periods, the learning agent interacts with the environment, selects various actions, receives related rewards, and updates the Q-values accordingly. This process of interaction and updating is repeated continuously. After a specified number of iterations, it is expected that the learning agent will have found the optimal settings for the SMPC protocol parameters, resulting in improved protocol performance in terms of efficiency and security metrics.

### C. Continuous Optimization

To ensure ongoing improvement, the agent continuously tests new settings and updates the Q-values to adapt to environmental changes and evolving requirements.

## VI. COMPARISON AND RESEARCH RESULTS

The use of model-free reinforcement learning to optimize SMPC protocols offers significant advantages in terms of continuous improvement, and balancing performance and security metrics. In contrast, traditional optimization methods may have limitations when faced with environmental changes and the discovery of optimal settings. A comparative analysis between RL and traditional methods in optimizing SMPC protocols reveals significant differences in key performance metrics. This analysis highlights the merits of RL regarding flexibility, optimization process, efficiency, and scalability.

SMPC protocols are essential for maintaining data privacy and security during collaborative computations. Optimizing these protocols is crucial for enhancing performance while maintaining robust security. This diagram provides a comparative analysis between RL-based and traditional optimization methods, focusing on four key features: optimization process, efficiency, and scalability.

TABLE I. COMPARISON OF USING RL VERSUS NOT USING RL IN OPTIMIZING SMPC PROTOCOLS

| Feature | RL (Reinforcement Learning) | Non-RL (Traditional Methods) |
| --- | --- | --- |
| Optimization Process | Finds optimal settings through exploration and exploitation | Relies on predetermined algorithms and methods which may not explore all possible settings |
| Efficiency | Effectively optimizes performance and security metrics | Faster and more resource-efficient but may not achieve full optimization |
| Scalability | Can adapt to increasing size and complexity of the problem | Scalability may be limited by the capabilities of traditional optimization methods |

In the diagram, the horizontal axis represents the features: optimization process, efficiency, and scalability. The vertical axis assigns scores from 0 to 10 for each feature. Blue bars indicate the scores for each feature using RL, while orange bars represent the scores without using RL.

RL excels in finding optimal settings through its exploration process. Unlike traditional methods, which rely on pre-determined algorithms, RL continuously learns and improves through interaction with its environment. This dynamic optimization process results in more effective and efficient protocol settings.

RL effectively balances performance and security metrics. While traditional methods may initially be faster and more resource-efficient, they often fall short in achieving complete optimization. RL, through its learning-based approach, improves the balance between computational cost and security, providing superior overall performance.

RL is well-equipped to manage increasing complexity and problem size. Traditional optimization methods may struggle with larger data sets and more complex computations, whereas RL can effectively handle these challenges.

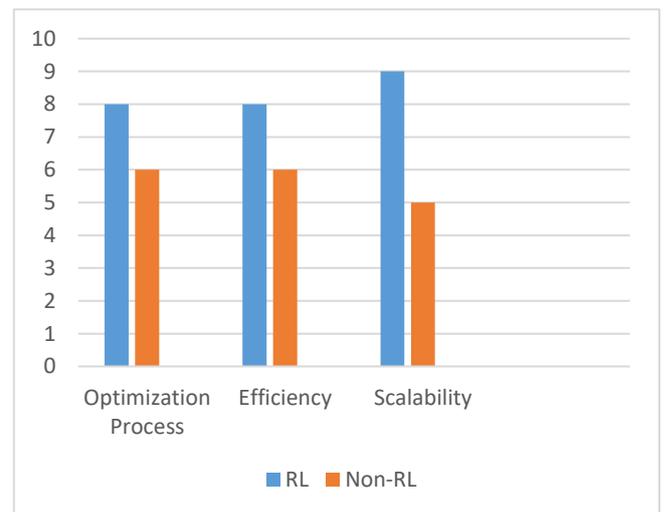

Fig. 4. *Empirical and Numerical Comparison Chart: Using RL vs. Not Using RL Method*

TABLE II. EMPIRICAL AND NUMERICAL COMPARISON OF USING AND NOT USING THE RL METHOD

| Feature | RL | Non-RL |
| --- | --- | --- |
| Optimization Process | 8 | 6 |
| Efficiency | 8 | 6 |
| Scalability | 9 | 5 |